\documentclass[aoas,preprint]{imsart}

\usepackage{graphicx,color}

\RequirePackage{amsthm,amsmath}
\RequirePackage{natbib}

\arxiv{math.PR/0000000}

\begin{document}

\begin{frontmatter}
\title{Modelling the impact of human activity on nitrogen dioxide concentrations in Europe.}
\runtitle{Modelling the impact of human activity on air pollution}

\begin{aug}
\author{\fnms{Gavin} \snm{Shaddick}\thanksref{m1}\ead[label=e1]{g.shaddick@bath.ac.uk}},
\author{\fnms{Haojie} \snm{Yan}\thanksref{m1}\ead[label=e2]{hy204@bath.ac.uk}}
\and
\author{\fnms{Danielle} \snm{Vienneau}\thanksref{m2}
\ead[label=e3]{d.vienneau@imperial.ac.uk}
}

\runauthor{G. Shaddick et al.}

\affiliation{University of Bath\thanksmark{m1} and Imperial College, London\thanksmark{m2}}

\address{Department of Mathematical Sciences, \\
University of Bath, UK. \\
\printead{e1}\\
\phantom{E-mail:\ }\printead*{e2}}

\address{Department of  Epidemiology and Biostatistics,\\ Imperial College, UK.\\
\printead{e3}\\
}
\end{aug}

\begin{abstract}
Ambient concentrations of many pollutants  are associated with emissions due to human activity, such as road transport and other combustion sources. In this paper we consider air pollution as a multi--level phenomenon within a Bayesian hierarchical  model. We examine  different scales of variation in pollution concentrations ranging from large scale transboundary effects to more localised effects  which are directly  related to human activity. Specifically, in the first stage of the model,  we  isolate underlying patterns in pollution concentrations due to global factors such as underlying climate and topography, which are modelled together with spatial structure.  At this stage measurements from monitoring sites located within rural areas are used which, as far as possible, are chosen to reflect background concentrations. Having isolated these global effects, in the second stage we assess the effects of human activity on pollution in urban areas. The proposed model was applied to concentrations of nitrogen dioxide measured throughout the EU for which significant increases  are found to be associated with human activity in urban areas.  The approach proposed here provides valuable information that could be used in performing health impact assessments and to inform policy.
\end{abstract}

\begin{keyword}
\kwd{Air pollution}
\kwd{Bayesian hierarchical models}
\kwd{spatial modelling}
\end{keyword}

\end{frontmatter}

\section{Introduction}

Modern research into, and management of, air pollution began in the middle of the twentieth century when serious concern arose about the possible effects of air pollution on health. To a large extent, this was driven by a series of high profile air pollution episodes, such as those in the Meuse River Valley, Belgium in 1930  \citep{heimann:61,ayres:72,pope:95,anderson:09} and 
    Donora Pennsylvania in 1948 \citep{anderson:67,snyder:94,chew:99}. 
In 1952 episodes of smog  in London  were associated with over 4000
deaths, resulting in the passing of the
Clean Air Act \citep{brimblecombe:87,giussani:94,brunekreef:02,stone:02}. In the U.S. problems of air pollution gradually rose together
with urbanization and led to the  first federal air pollution legislation in 1955. 
Early air pollution control legislation was focused on setting restrictions on the use of smoke-producing
fuels and smoke-producing equipment \citep{Garner:69,Stern:73}. More recently, air quality standards such those issued by the  WHO   relate to a specific pollutants, such as particulate matter (PM), ozone (O$_3$), sulphur dioxide (SO$_2$), carbon dioxide (CO) and nitrogen dioxide (NO$_2$)   \citep{who2005air}. \\

Despite decreasing levels of air pollution since regulation, many epidemiological studies have reported associations between air pollution and adverse health outcomes at relatively low levels.  The majority of studies have shown relationships between short-term effects of air pollution and health and recently there have been a number of large multi-city studies including  Air Pollution and Health: A European Approach (APHEA I and II, \cite{katsouyanni1997short, katsouyanni2001confounding}) in Europe and 
the National Morbidity, Mortality and Air Pollution Study (NMMAPS, \cite{dominici2002air}) in the U.S.
A smaller  number of studies have  investigated possible longer-term effects, including \cite{Abbey:99,Hoek:02,Nafstad:03,Finkelstein:03,Jerrett:05,Rosenlund:06,elliott:07}. More recent  quality standards, for pollutants such as    PM, O$_3$ and NO$_2$  are specifically intended to protect the public from the possible health effects of pollution   \citep{who2005air}. \\

The term  air pollution in its general form represents a complex mixture of many different components with  individual  pollutants classified as either primary or secondary. Primary pollutants are
those emitted directly from a source, whereas secondary pollutants are formed in
the atmosphere through chemical reactions. Ambient concentrations of many pollutants, for example NO$_2$ and CO,  are associated with human activity, such as road transport and other combustion sources, and would be expected to be higher in urban areas. Conversely,  ozone is
almost entirely a secondary pollutant, but is subject to scavenging
by nitrogen oxides, so tends to reach its highest concentrations in
rural areas remote from major traffic sources. \\

When modelling concentrations of air pollution, 
it is useful to recognise three components of variation in the monitored concentrations, operating at different spatial scales. In most cases we would expect to find some degree of broad--scale variation or ÔtrendÕ that can perhaps be represented by a relatively simple, global surface. Superimposed on this we would expect to find more local variation, associated perhaps with the distribution of emission sources and the effects of local topography or land cover. At an even more local level, we can expect to find short-range variation (e.g. from one side of a street to another) which is probably beyond the resolution of the data considered here, but which may occur as ÔnoiseÕ in the monitored data. Measurement errors, differences in monitoring methods and in sampling times may also contribute to this noise.  \\

In this paper,  we aim to investigate these different components of variation within levels of NO$_2$ throughout the EU. The approach essentially comprises two stages, firstly we attempt to identify monitoring sites in rural locations which, as far as possible, might be expected to reflect background levels of pollution and to use these to  isolate underlying global effects. In all but the most remote of locations there will still be  emissions due to   human activity which needs to be acknowledged when trying to estimate the effects of topography and climate. This is achieved by using covariate information based on land--use, roads and population density as proxies to represent the intensity of human activities within the first stage of a Bayesian hierarchical model which also incorporates spatial structure. \\

At the global scale, we use altitude and the distance from the sea  which have been shown to be associated with levels of NO$_2$ \citep{briggs:05, madsen:07, ross:05, hoek:08} together with meteorological factors such as temperature and wind. It is assumed that although  human activity may affect some factors such as local temperature  it  will have little effect on global climate  over a wider region, e.g.  annual average temperature for both rural and urban areas is still dominated by climate. When modelling local variation, we consider covariates  such as  traffic density and population, which have been shown to have strong relationships with NO$_2$ \citep{briggs:97, henderson:07, gilbert:05,briggs:00, carr:02, briggs:05, ross:05}. \\

The second stage of the process is to assess the affects of human activity  within urban areas. Using estimates of the global effects from the first stage together with the spatial structure, we make predictions at the locations of monitoring sites in urban areas based purely on their topography and climate, i.e. as if there was no human activity. By comparing these predictions with the observed concentrations, we aim to identify to which levels of NO$_2$ can be attributed to urban human activity as represented by a set of urban level covariates. \\

The remainder of this paper is as follows, in Section 2 we give details of  NO$_2$ concentrations measured at background and urban locations within Europe, Section 3 provides details of the structure of the Bayesian hierarchical model and Section 4 presents the results of applying the models.  Finally,  Section 5 contains a discussion and details of potential future developments.

\section{Data}

\begin{figure} \centering \fbox{\includegraphics[scale=0.4]{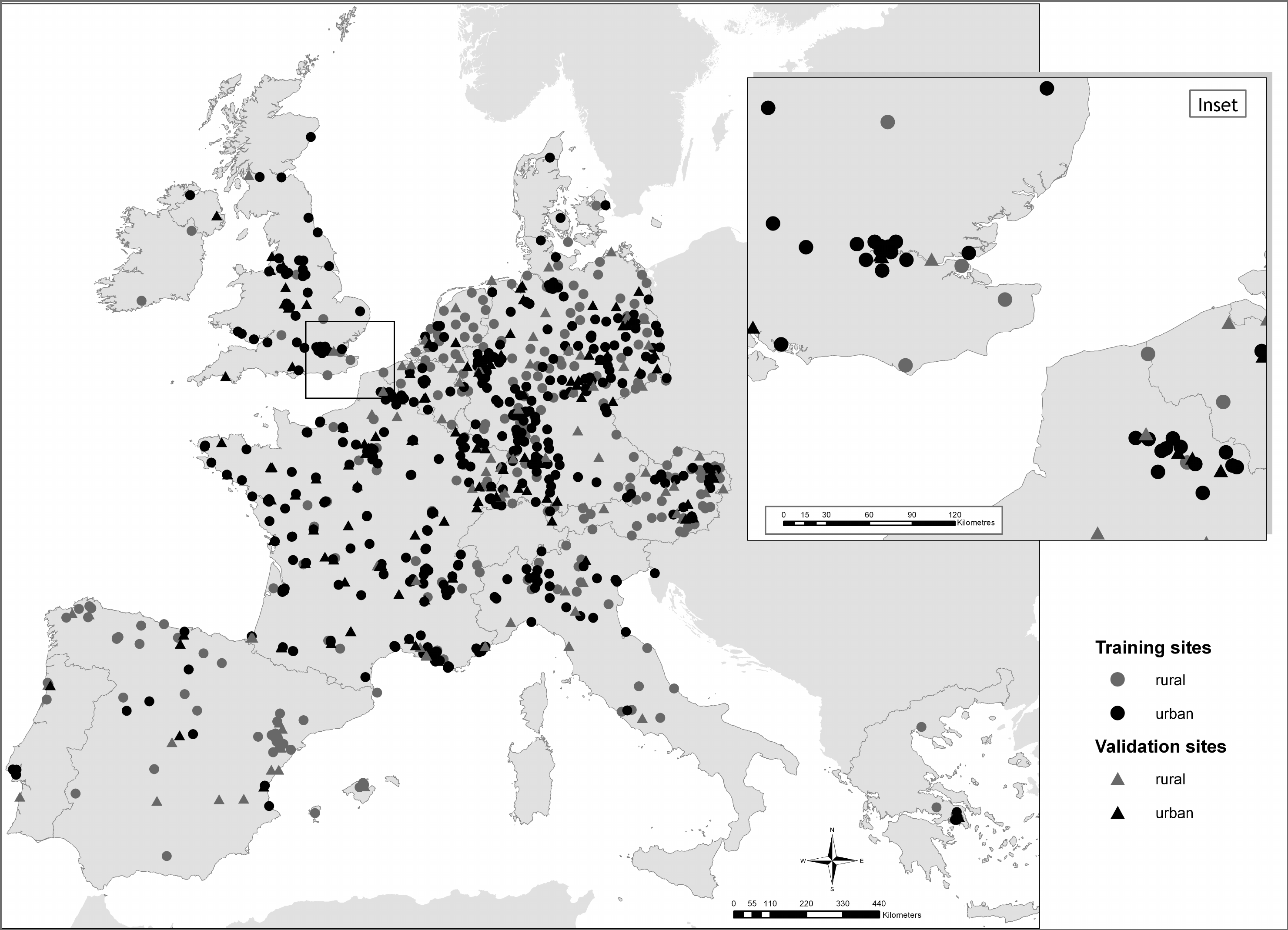}} \caption{\label{Fig1} Locations of nitrogen dioxide monitoring sites at  rural   (circles) and urban (triangles)  locations. }
\end{figure}

The study area comprises the EU-15 countries; Austria, Belgium, Denmark, France, Germany, Greece, Ireland, Italy, Luxembourg, The Netherlands, Portugal, Spain and the United Kingdom. However Finland and Sweden are excluded due to lack of data. Annual averages of NO$_2$  for 934 background monitoring  sites in 2001 with $\geq$75\% data capture  were extracted from the Airbase database  (www.eea.europa.eu/themes/air/airbase). Monitoring sites are distinguished according to site type; traffic, industrial and background and station location (urban, suburban and rural).
At the time of study, these classifications were found to be incomplete and inconsistent across countries. To address this a contextually based classification,  derived on the basis of discriminant analysis with consistent EU-wide land cover, was used to identify background monitoring sites. These background sites were further classified as either rural or urban \citep{vienneau2009gis}.  This GIS based contextual rural/urban classification also enabled classification of areas (1 km cells) across the study area, which is necessary for prediction and mapping purposes.  The set of background monitoring sites were randomly allocated to either a training  or validation set (comprising 75\% and 25\% of sites respectively), stratified by rural/urban status and country . The training and validation datasets comprise 250 rural, 458 urban and 86 rural, 140 urban sites respectively.  The locations of the rural and  urban training sites can be seen in Figure \ref{Fig1}. \\

 \begin{center} INSERT FIGURE 1 HERE \end{center}

\begin{figure}
\begin{center}
\includegraphics[scale=0.65]{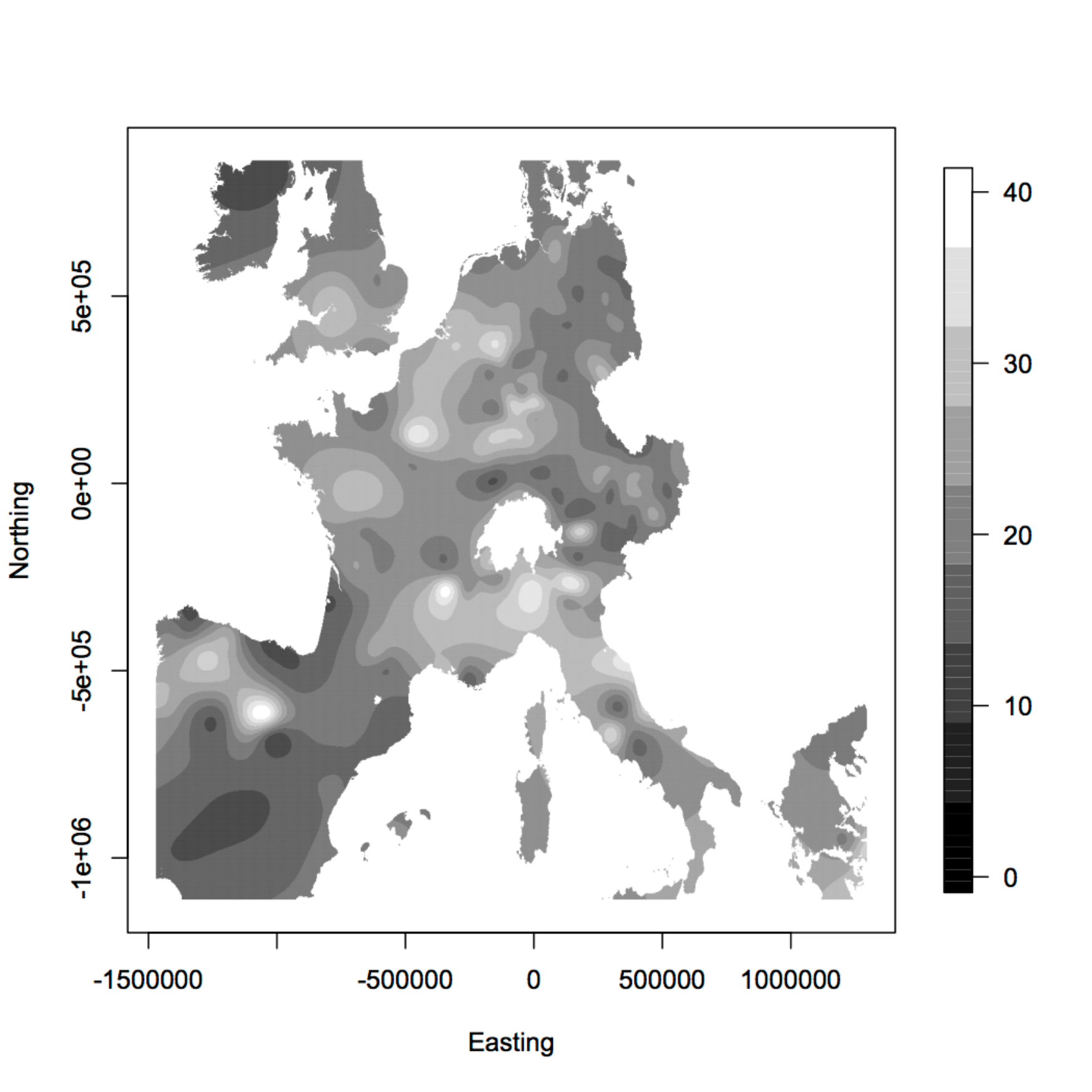}
\caption{Smoothed plot of concentrations of NO$_2$ at background sites in rural locations. } \label{smoothplot}
\end{center}
\end{figure}

 A summary of the concentrations from the different locations can be seen in  Table \ref{ch3:table:no2:pollutant:summary}. As might be expected, the levels recorded at urban locations are higher than those at background locations with more variability being observed within the urban locations.  Figure \ref{smoothplot} shows the concentrations of NO$_2$ at the background monitoring sites located in rural areas, smoothed using multi-level B-splines \cite{lee1997scattered}.  Although these sites were chosen to ideally reflect background concentrations the effects of human activity are clearly observable particularly when rural areas are in close proximity to large cities.
 
 \begin{center} INSERT TABLE 1 HERE \end{center}

 \begin{center} INSERT FIGURE 2 HERE \end{center}

Covariate data were obtained from a number of sources, including CORINE (land cover),  TOPO30 (topographical information), AND (transport networks), MARS (meteorology) and SIRE  (population) databases and was compiled on a 1 km grid.  The geographical information system (GIS) database is fully detailed in \cite{robbb} and briefly summarised here. Covariates were computed at different spatial scales  with the aim of representing different scales of variation: local (the immediate 1 km square within which the monitoring site lies), zonal (within the surrounding 5km neighbourhood) and regional (within the surrounding 21 km area). In each case, covariates were computed by defining a circular window around the centre of the target grid cell, and calculating the area-weighted total or average for that measure within the window. In the case of roads the value is the total length within the area and for land--use variables it is the percentage of the area attributed to that use.  In the modelling it is assumed that  there is a linear relationship between covariates and air pollution concentrations and so transformations were considered. For both altitude and distance to sea, the following transformation was used to address non-linear relationships; $\sqrt{x'/max(x')}$, where $x' = x-min(x)$.\\

Here, the covariates are classified into three groups; global, rural and urban.Global level variables are those based on climate and topography and include altitude, distance to sea and meteorological variables;  seasonal temperatures, wind speed, days of calm and annual radiation (9 variables). Due to the high
 levels of collinearity observed in these climate variables, principal component
 analysis was used to produce five factors, which accounted for 97\% of the total
 variation.  These five climate factors represent areas which (1) are hot year round and windy, (2) have hot summers, cooler winters, (3) are cool year round,  wet and calm, (4) are cool year round, dry and calm and (5)  have cold calm winters and warm windy summers.  Rural and urban level covariates are based on land--use, roads and population density and   are used as proxies to represent the intensity of human activities.  \\
 
  \begin{center} INSERT TABLE 2 HERE \end{center}
 
 Table \ref{LU data} gives the three sets of covariates; global, rural and urban together with the scale at which they are calculated and the mean levels for rural and urban sites. Of the global variables, which will be used in both the models for rural and urban sites,  it can be seen that there is little difference between rural and urban monitoring sites in terms of the climate variables  and distance to sea but that the altitude of the rural  sites    are on average over twice that of the urban ones.  Rural and urban variables are used in modelling concentrations at rural and urban sites respectively.  \\

  \begin{center} INSERT FIGURE 3 HERE \end{center}

\begin{figure}
\begin{center}
\includegraphics[scale=0.6]{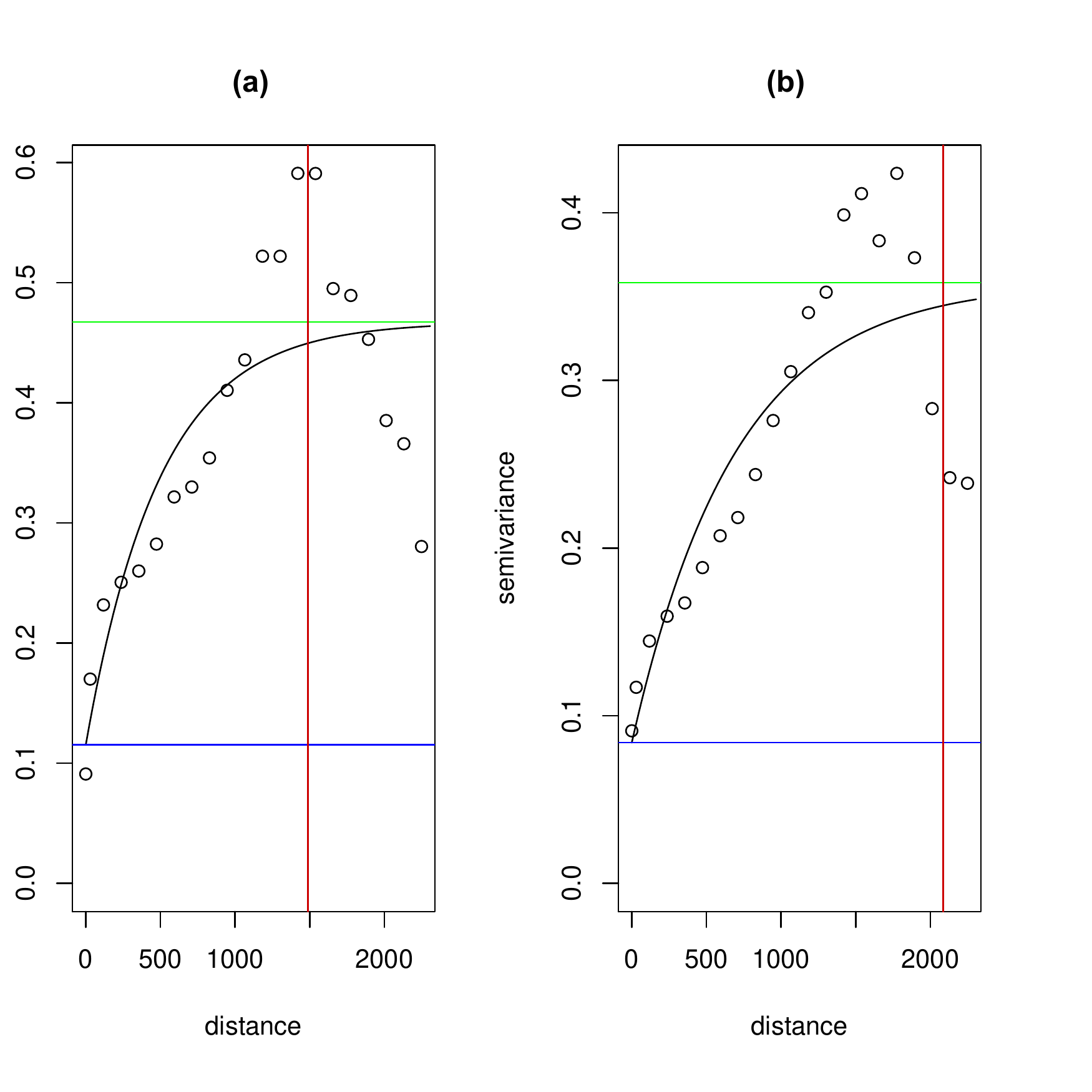}
\caption{Variogram for (a) log values of Nitrogen Dioxide (NO$_2$); (b) residuals after fitting model with global and rural covariates. Lines show fitted exponential curve (black), nugget (blue), partial sill (green) and effective range at which correlation falls to 0.05 (red).} \label{ch4:no2:variogram:EU1}
\end{center}
\end{figure}

The plot of concentrations at rural locations presented in Figure \ref{smoothplot} indicates the presence of spatial auto--correlation. 
Figure \ref{ch4:no2:variogram:EU1} shows the empirical variogram for measurements on the log scale from the rural sites (in the left panel) and for  residuals from  a multiple linear regression model using global covariates (right panel). Evidence of  spatial correlation is apparent from this figure,  the variogram increases from approximately 0.1 up to 0.45 corresponding to a strong correlation between close locations and then variogram levels off from 1500 km as the correlation decays to zero. The decrease of the nugget (from above 0.1 to less than 0.1) and maximum value  (from ca. 4.5 to ca. 0.36) of the variogram in the right panel indicates the introduction of covariates reduces the overall spatial variation in the residuals (compared with what is essentially a model with just an intercept term). From these figures, it is suggested that there is is evidence of spatial structure in the data which should be incorporated in the model.\\

\section{Bayesian hierarchical model}

The Bayesian hierarchical model developed here has three main levels; (i) a global model which   relates  concentrations at rural monitoring sites to sets of global and rural covariates  together with residual spatial structure, (ii) prediction using the global and spatial effects at urban locations and (iii) estimation of the effect of urban covariates using the subsequent differences in predicted and observed concentrations.  In addition, a fourth level defines the hyperpriors which are required for any Bayesian analysis.\\

\cite{ott1990} has suggested that 
a log transformation is  appropriate for modelling 
 pollution concentrations, because in addition to the desirable properties of
right-skew and non-negativity, there is justification in terms of
the physical explanation of atmospheric chemistry and so the logs of the annual means at the monitoring locations are used throughout, with transformation back to the original scale for the presentation of a selection of the results. \\

\subsection{Stage one: global level model}
The aim of this stage of the model is to estimate the global effects  which are then used to predict at urban locations in the next stage, allowing for the effect of the rural covariates. \\

Let $Y_{i}$ represent the log transformation of the annual average NO$_2$ concentration measured at rural sites, $i$,

 \begin{eqnarray}
\label{stage1}
 Y_{i}=\beta_{0}+\sum^{G}_{k=1}\beta_k X_{i}+\sum^{p}_{k=G+1}\beta_k X_{i}+m_{i}+\nu_{i}
 \end{eqnarray}

where  $i=1,...n$. The overall mean is denoted by $\beta_0$ and the global and rural covariates at rural locations by the  $n \times p$ matrix $X$ which is partitioned into $(X^G, X^R)$ denoting the G global covariates and R rural covariates. The associated regression parameters are  $\beta_1,...,\beta_p$. The random error terms,  $\nu_i$, are assumed i.i.d. $N(0,\sigma^2_{\nu })$ with $m_i$. A  set of spatial effects, $\mathbf{m} = (m_1, ..., m_n)$ are  assumed to arise from the multivariate normal distribution,  $MVN(0_S,\sigma^2_{m} \Sigma_{m})$,
where $0_S$ is an $S \times 1$ vector of zeros, $\sigma^2_{m}$ the between-site variance and $\sum_m$ is
the $S \times S$ correlation matrix. The  correlation between sites is related to the distance between them and takes    the form  $ f(d_{i,j},\phi) = \exp{(-\phi d_{i,j})}$
where $\phi>0$ describes the strength of the correlation--distance relationship, which results in a isotropic and stationary spatial model. 

\subsection{Stage two: prediction at urban locations}
\label{ch4:sec:pred}

In this fully Bayesian framework, estimation of the covariate effects in the global level model and prediction at the urban locations is performed simultaneously. The uncertainty in estimating the coefficients of the global model is therefore  acknowledged and `fed through' the model to the predictions (this stage) and further to the estimation of the coefficients in the urban model (stage 3). However it is noted that feedback is `cut' between the third and second stages  \citep{spiegelhalter:98}.  It  is not intended that the urban sites should inform the estimation of the global effects which should be based on data from the rural sites which are intended to provide information on background concentrations.\\

  If the random error terms, $\nu_j$, in (\ref{stage1}) are uncorrelated, then a prediction at a new location, $j$ will take the form

 \begin{eqnarray}
 \label{stage2}
 \hat{Y}_{j}=\hat{\beta}_{0}+\sum^{G}_{k=1}\hat{\beta}_k X_{j}+ \hat{m}_{s^{'}}
 \end{eqnarray}

This can be viewed as two separate process; the first predicting covariate effects at new locations, using values of the global covariates at the urban locations with the values of the rural covariates which are related to human activities set to zero,      and the second predicting the spatial effect.  The spatial component is calculated using properties of the multivariate normal
distribution. If $\mathbf{m}=(m_1,..,m_n)'$ are the observed values at the monitoring locations, then the conditional distribution of $m_{j}|\mathbf{m}$ at a new location, $j$, will be
normally distributed with mean and variance given by
\begin{eqnarray}
E[m_{j}|\mathbf{m}] = \sigma^{-2}_m\mathbf{\delta}_{j}'\Sigma^{-1}_m \mathbf{m},
\end{eqnarray} and
\begin{eqnarray} var(m_{j}|\mathbf{m}) = \sigma_m^2(1 - \mathbf{\delta}_j' \Sigma^{-1}_m
\mathbf{\delta}_j),
\end{eqnarray} respectively, where $\delta_j$ is the vector of distances between the new location and the monitoring sites and $\delta_{ij} = f(d_{ij}, \phi)$.

\subsection{Stage three: estimating urban effects}

The residuals from the predictions using the global model at the urban locations are then regressed against the urban level covariates with   the uncertainty from the previous stage being  propagated through the model.
 \begin{eqnarray}
 \label{stage3}
 (Z_j-\hat{Y}_{j})=\gamma_{0}+\sum^{q}_{l=1}\gamma_k W_{j}+\omega_j,
 \end{eqnarray}

where $Z_j$ is the log transformation of the annual average  measured at  urban locations, $j=1,...,m$, $\gamma_1$ represents the overall difference between the predicted and observed levels. Urban covariates are denoted by  the  $m \times q$ matrix $W$ with  associated regression parameters,  $\gamma_1,...,\gamma_p$ and $\omega_j$ are random error terms at  urban locations  which are assumed i.i.d.  $N(0,\sigma^2_{\omega})$.

\subsection{Stage four: hyperpriors}

Prior distributions were assigned to all random variables, e.g. covariate effects, site
effects and variances. Vague normal priors are assumed for the intercept and covariate terms, $\beta_0, \gamma_0, \beta_j$ and $\gamma_k \sim N(0,1000)$ with the precisions of the error terms $\nu^{-2}, \omega^{-2}$ assumed to be Gamma distributed, $Ga(1,0.01)$. A uniform prior is used for the strength of the correlation--distance relationship with the limits of $\phi$ being based on beliefs about the relationship between correlation and distance. For example, the distance, $d$, at which
the correlation, $\rho$, between two sites might be expected to fall to a particular level would be $d = -\log(\rho)/\phi$.  Vague normal (as above) prior distributions are also assigned to the  predictions at the urban locations which are in essence treated  as unknown parameters with  inference on the parameters of
interest, i.e. the urban coefficients, being performed via averaging over the distributions of
these predictions.

\subsection{Inference}
\label{ch4:ModelInference}

The joint distribution of the parameters is:

 \begin{eqnarray} \nonumber
 p(\beta, \gamma, m,\phi, \sigma^2_{\nu},\sigma^2_{\omega}\sigma^2_{m}|y, z) 
 &\propto& p(y|\beta, m, \sigma^2_{\nu}) \\ \nonumber
 &&   p(z|\gamma,  \sigma^2_{\omega}, \beta, m) \\ \nonumber
 && p(m|\phi,\sigma^2_{m}) \\
 && p(\gamma)p(\beta)p(\sigma^2_{\nu})p(\phi) p(\sigma^2_{m})
 \end{eqnarray}

which is analytically intractable but samples from this
distribution may be generated in a straightforward fashion using
Markov Chain Monte Carlo (MCMC) \citep{smith:roberts:93}.  The prior distribution for $\beta$, $\sigma^2_{\nu}$ and $\sigma^2_{m}$ are chosen to be conjugate and Gibbs sampling can be used for all parameters, with the exception being $\phi$ which has a uniform prior and thus the full conditional is not available in closed form, requiring a Metropolis--within--Gibbs step. This
was performed using the WinBUGS software
(\cite{spiegelhalter:98}).
\\

\section{Results}

For each of the models presented two  MCMC chains were run (for each parameter) with a minimum of 40,000 iterations as \emph{burn-in}  and at least a further 10,000 samples per chain used to calculate summaries of the posterior distributions. Convergence was assessed both visually and by use of the Gelman-Rubin statistic \citep{gelmanrobin}, which measures the ratio of the between and within chain  variances. All parameters  achieved convergence, although it is noted that the spatial correlation--distance parameter, $\phi$, generally took longer to converge that the other parameters.\\

 In fitting the models vague normal priors $N(0,0.001)$ were assigned to the  covariate effects and intercept terms, while for the precisions of the random error and spatial terms 
$Ga(1,0.01)$ were assumed.  For the distance--correlation parameter, $\phi$, in the global model
the limits were chosen to
represent a drop to 0.01 at 25km and 2000km, i.e. representing  strong and weak decays in  correlation over distance respectively.\\

\subsection{Isolating global effects}

Table \ref{ch4:result:rural:global} gives the results of fitting  models to data from  the rural sites
The most significant effect was a decrease in levels of NO$_2$ with   increasing altitude and  (in the model with both global and rural covariates) a significant positive effect was observed in relation to  the fifth climate factor. 
When comparing a model with global level covariates with one without covariates, there is a reduction in the spatial variance, $\sigma^2_m$ indicating  that much of the spatial variation in global NO$_2$ can be explained by the covariates, leaving less unexplained variation to be `mopped up' by the spatial residual term as indicated by the variograms (Figure \ref{ch4:no2:variogram:EU1}).   Their inclusion also   results in a reduction in the decay of  correlation over distance, meaning that correlations will be greater once covariate effects have been accounted for. As an example,  the correlation at 100km is 0.01 without covariates and 0.44 when they are included. \\

  \begin{center} INSERT TABLE 3 HERE \end{center}

The covariates also improve the model's ability to predict at the validation locations. Calculating summary functions such as  the root mean squared error (RMSE)  at each iteration of MCMC simulation  results in a posterior distribution as it is a simple function of other parameters being estimated. In this case,   the median of the posterior distribution of the RMSE reduces from 12.9 to 9.5  when global and rural covariates are added to the model with a corresponding increase in R$^{2}$ from 17.5 to 44.8\%. Of the 86 rural validation sites  the vast majority (86\%, 74/86) of the observed values lie within the 95\% credible intervals for the predictions. Again this is  an improvement over a model with just global covariates where the corresponding value is 58\%, indicating that there is a component of levels of pollution at rural background locations which is still related to emissions from human activity which needs to be accounted for before the global effects can be examined.  \\

\begin{figure}[hptb]
\centering
\includegraphics[scale=0.4]{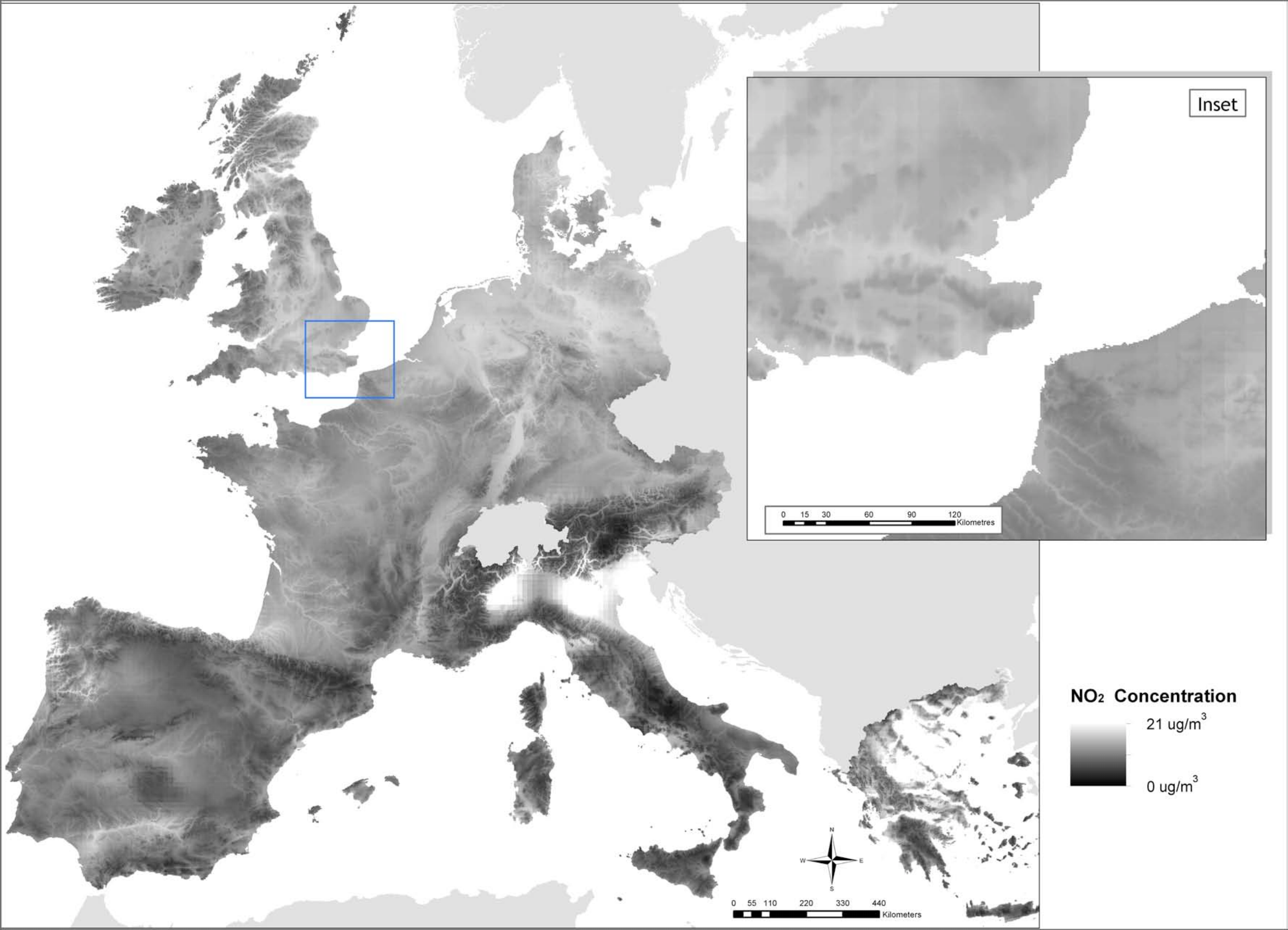}
\caption{Predicted concentrations of NO$_2$ using global effects. \label{globalmap}}
\end{figure}

Figure \ref{globalmap} shows  predicted concentrations of NO$_2$ throughout the EU using estimates of the global effects, i.e. based purely on their topography and climate.  In comparison with Figure \ref{smoothplot}  which showed the measured concentrations, there is an overall reduction in levels of NO$_2$ with the effect of the urban areas close to the rural monitoring sites  being markedly reduced.

  \begin{center} INSERT FIGURE 4 HERE \end{center}

\subsection{Assessing the effects of human activity}

\begin{figure}[hptb]
\centering
\includegraphics[scale=0.6]{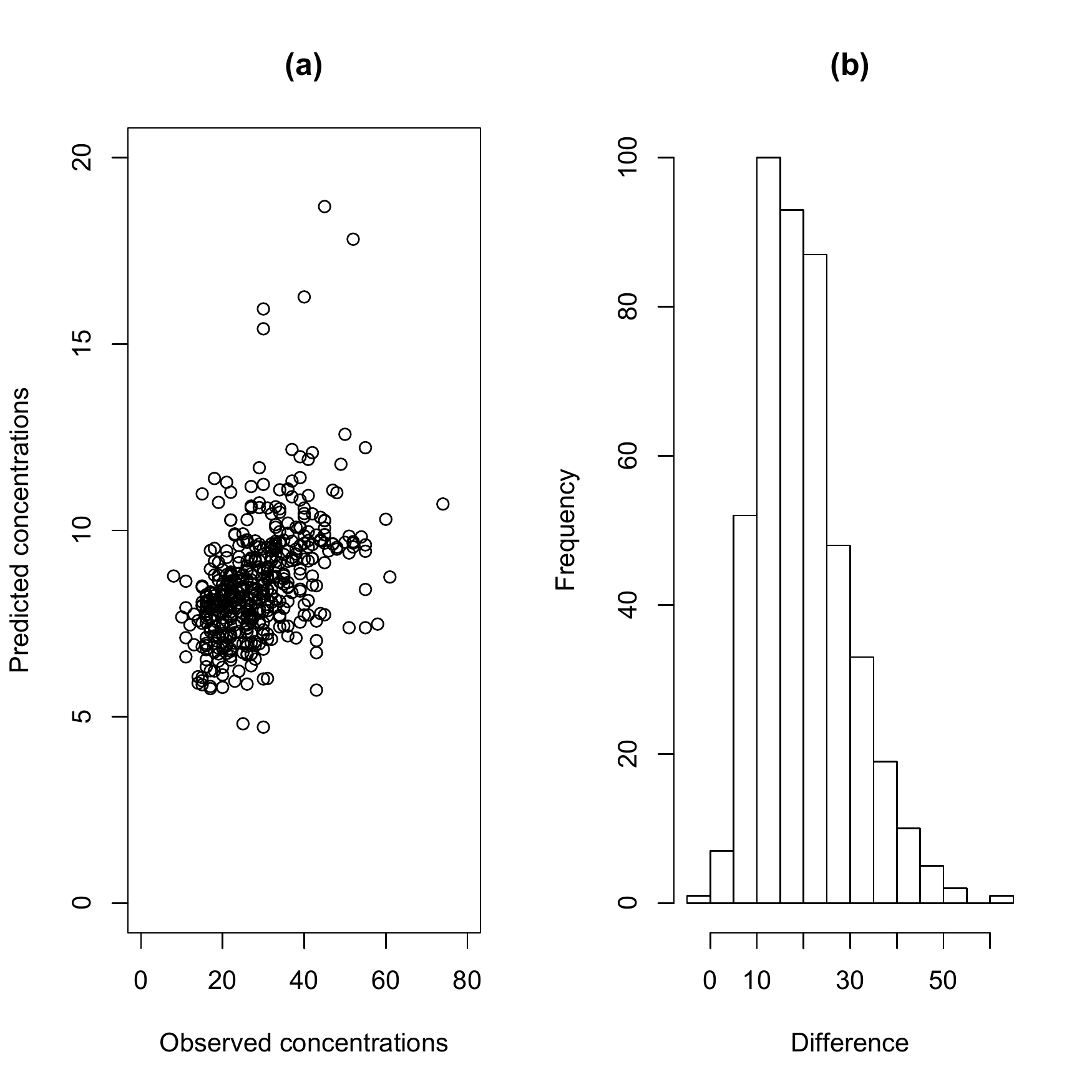}
\caption{Comparison of the observed concentrations at the urban locations with predictions from model using only global effects. Left hand panel (a) shows plot of predicted versus observed concentrations, right hand panel (b) shows a histogram of the differences.\label{UrbanComp}}
\end{figure}

The results of predicting at the urban locations using just the global and spatial effects can be seen in Figure \ref{UrbanComp}. The predictions are, as expected, much lower than the observed values with the median difference being 18.7 $\mu gm^{-3}$ (IQR, 13.1 - 25.4) and this difference will be examined in relation to urban factors.

  \begin{center} INSERT FIGURE 5 HERE \end{center}

Table \ref{ch4:result:model2} gives the results of fitting  models to data from  the urban sites, where differences between observed urban NO$_2$ concentration and that predicted as though they were background locations are explained by the effects of urban covariates. In a model without covariates the intercept term, representing the urban increment, is 3.294 representing an (significant) overall increase of 27 $\mu$g$m^{-3}$. In the model including urban covariates,   there is again a positive (significant) urban intercept term with all the urban level covariates show further positive associations with levels of NO$_2$, except for urban greenery.  Major roads has the largest significant effect with an relative increase of 1.06 $\mu$gm$^{-3}$ (=exp(0.0623)) associated with an increase of 1 km of road length (in the surrounding 1 km area) with high density residential also having a significant association. \\

  \begin{center} INSERT TABLE 4 HERE \end{center}

\section{Discussion}

In this paper we consider air pollution as a multi--level phenomenon within a Bayesian hierarchical model. Different scales of variation are considered ranging from large scale transboundary effects to more localised effects which are related to human activity. The aim of the first stage of the model is to isolate underlying patterns in pollution concentrations due to global factors, such as underlying climate and topography,  from those arising from land use and traffic. At this stage monitoring sites located within rural areas were used which, as much as is possible, were chosen to reflect background concentrations. However, in all but the most remote of areas there will still be some effect of human activity on levels of pollution and so carefully selected covariates representing
 emission sources, such as land cover or road density, were used at either zonal or  regional levels (representing the surrounding 5km and 21km respectively) to isolate global effects together with long-range spatial structure. \\
 
 Having isolated these effects, in the second stage of the model we assess the effects of human activity on levels of pollution in urban areas where such activity will be greatest.  We found a significant increase in levels of NO$_2$ in urban areas compared to that which might be expected based on global effects alone. The estimated increase from the second stage of the model was 27.0 $\mu$gm$^{-3}$ (95\% CI  26.1 -- 27.9) which  is considerably greater than the difference observed between the means of the concentrations observed at rural and urban sites (13  $\mu$gm$^{-3}$). This is because the concentrations observed at the majority of rural locations will inevitably include some component which is due to human activity in the surrounding area. They therefore cannot be assumed to give a true reflection of background concentrations without accounting for the resulting emissions as we have attempted to do here in the first stage of the model.  \\

 We assume that pollution varies smoothly on a global scale and that urban areas are embedded in an overall spatial surface. Here, this global spatial structure in the  first stage of the model is assumed to be isotropic and stationary. The assumption of isotropy can to some extent be assessed by constructing variograms in different directions, which in this case showed little evidence of ansitropy (not shown). The assumption of stationarity  might be tenable for  the  global scale, after adjustment for covariate effects, whereas using a such a model to address finer scale variation over a large number of urban areas  might be less reasonable.  The ability of a spatial model to  provide accurate predictions can to some extent be assessed by validation,  but there may be underlying problems based on the availability of data over the entire study area. For example, (i) the covariates may not fully reflect the areas in which the pollution is measured, i.e. other important covariates may have been excluded or are not available; (ii) the spatial structure in the model is sufficiently flexible to be able to accurately reflect complexity of air pollution process and (iii) the location of monitoring sites may  not fully represent  the spatial pattern of pollution over the study region, i.e. the monitoring sites are unevenly distributed over the region and are not able to represent the underlying spatial process.  \\
 
   The approach used here is similar in concept to the two step regression modelling strategy of \cite{robbb}, although the specific aim of that paper was to perform mapping. Their two step procedure involved fitting two separate  models  and using the prediction from the first as a fixed covariate for the second, thus ignoring  the fact that the prediction is an estimate based on the first regression and is thus subject to uncertainty. By performing  both models simultaneously within a Bayesian hierarchical framework, this uncertainty is acknowledged and correctly `fed through' the model. We performed a comparative analysis using a two-stage approach in which uncertainty was not fed though the model and found the confidence intervals were  much narrower. For example, in the case of the estimate of the effect of major roads the width of the credible interval reduced from 0.115 to 0.056 (data not shown). \\
 
An alternative approach would have been to combine the rural and urban sites in a single model, however that would lead to high--levels of collinearity between measurements at different scales, or to fit global and urban effects together using data from urban locations.  However,  the influence of human activity on concentrations of pollution in urban locations is so strong that, even after covariate adjustment, they cannot be used to represent  background levels. In practice, this means  that the urban covariates are likely to   dominate the global ones to such an  extent that interpreting the  global part of the model would be very  difficult. In performing such an analysis we found, for example, that  the effect of altitude was positive which is entirely counter intuitive for NO$_2$.   In contrast, the approach used in this paper allows us to combine data from both rural and  urban sites, to estimate background concentrations and  therefore quantify the contribution to air pollution attributable to human activity within a coherent modelling framework. \\

The models proposed here provide valuable information that could be used in performing health impact assessments and to inform policy. For example, further  research could utilise the the differences in urban and background concentrations in order to  assess the health risk of  air pollution that is attributable to human activity.

\bibliography{hybib_final2}
\bibliographystyle{chicago}

\newpage


\begin{table}
\caption{Summary of NO$_2$ concentrations by site location; annual means for 2001}
\label{ch3:table:no2:pollutant:summary}
\begin{center}\scalebox{1}{
\begin{tabular}{ccccccccc}

   Location        	&	       Mean 	&	SD	&	Median	&	IQR	&	Min-Max	\\
\hline						 					
All  	&	24	&	12	&	23	&	(16-31)	&	2-74	\\
						 					
Rural 	&	16	&	9	&	14	&	(9-21)	&	2-43	\\
											
Urban  	&	29	&	10	&	27	&	(21-35)	&	8-74	\\
\hline
\end{tabular}}
\end{center}
\end{table}

\begin{table}
\caption{Summary (means) of  covariates at locations of NO$_2$ monitoring sites (training set) at all, rural and urban locations. See text for descriptions of the covariates and the resolution at which they are computed.  \label{LU data}}
\begin{center}
\begin{tabular}{lccc} 
Covariate & All sites & Backgound & Urban \\ \hline
{\em Global} &&&\\
Altitude (m)&220&360&145\\
Distance to sea (m)&202&198&205\\
Climate factor 1 &0.83&0.71&0.90\\
Climate factor 2 &-0.42&-0.20&-0.55\\
Climate factor 3 &0.25&0.26&0.24\\
Climate factor 4 &0.05&-0.02&0.08\\
Climate factor 5 &-0.03&-0.03&-0.03\\ \\
{\em Rural}&&&\\
Major roads (5 km)&10.85 &0.65&14.64\\
Minor roads (5 km)&35.59&2.42&42.70\\
High density residential (5 km) &5.74& 0.57 & 8.56\\
Low density residential (5 km) &26.14   &6.05 &37.10 \\
Agriculture (5 km) &45.95& 50.5 &44.60\\
Non-rural built up (21 km) & 3.72&1.30 &5.04\\
Forestry (21 km) &19.84&27.05 &15.90\\ \\

{\em Urban}&&&\\
Major roads (1km)&0.65&0.25&0.87\\
Minor roads (1km)&2.42&1.32&3.02\\
High density residential (1 km)&11.00&0.64&16.70\\
Low density residential (1 km)&38.8&7.54&55.9\\
Industry (1 km)&6.12&1.08&8.86\\
Transport (1 km)&0.93&0.05&1.41\\
Sea port (1 km)&0.30&0.03&0.48\\
Air port (1 km)&0.20&0.07&0.45\\
Construction (1 km)&0.56&0.36&0.67\\
Urban Greenery (1 km)&2.24&0.18&3.36\\

\hline
\end{tabular}
\vspace{3mm}
\end{center}

\end{table}


\begin{table}[hptb]
\caption{Rural monitoring sites: Summaries of posterior distributions of parameters; medians and 95\% credible intervals for models with and without covariates. Models are fit with (i)  intercept term, (ii)  global covariates and (iii) global and rural covariates.}
\label{ch4:result:rural:global}
\begin{center}
\scalebox{0.75}{
\begin{tabular}{l|rrr|rrr|rrr}

 & \multicolumn{3}{c}{(i)} & \multicolumn{3}{c}{(ii)}  & \multicolumn{3}{c}{(iii)} \\
&     Median &   $2.5\%$ &  $97.5\%$   &     Median &   $2.5\%$ &  $97.5\%$
           &     Median &   $2.5\%$ &  $97.5\%$\\ \hline
           
Intercept	&	2.5830	&	2.4660	&	3.1790	&	3.2110	&	2.6640	&	3.8160	&	2.4150	&	1.8440	&	3.1000	\\
Altitude	&		&		&		&	-3.5580	&	-4.3720	&	-2.8000	&	-1.9970	&	-2.9720	&	-1.0870	\\	
Dist. sea	&		&		&		&	0.6367	&	-0.1733	&	1.3480	&	0.2754	&	-0.4621	&	0.8936	\\	
Climate factor 1	&		&		&		&	-0.1725	&	-0.3591	&	-0.0147	&	-0.1159	&	-0.2991	&	0.0283	\\	
Climate factor 2	&		&		&		&	0.0296	&	-0.0933	&	0.1023	&	0.0140	&	-0.0664	&	0.0771	\\	
Climate factor 3	&		&		&		&	-0.0930	&	-0.2523	&	0.0850	&	-0.0475	&	-0.2309	&	0.0942	\\	
Climate factor 4	&		&		&		&	-0.1839	&	-0.3424	&	-0.0294	&	-0.1213	&	-0.2744	&	0.0185	\\	
Climate factor 5	&		&		&		&	0.3096	&	-0.0190	&	0.6182	&	0.2880	&	0.0090	&	0.5467	\\	
	\hline
Major road 	&		&		&		&		&		&		&	0.0181	&	0.0031	&	0.0329	\\	
Minor road	&		&		&		&		&		&		&	0.0082	&	0.0012	&	0.0153	\\	
High density res. 	&		&		&		&		&		&		&	0.1925	&	0.0221	&	0.3666	\\	
Low density res.	&		&		&		&		&		&		&	0.0070	&	-0.0024	&	0.0164	\\	
Agriculture	&		&		&		&		&		&		&	0.0039	&	0.0009	&	0.0069	\\	
Non-rural built up	&		&		&		&		&		&		&	0.0383	&	0.0071	&	0.0690	\\	
Forestry	&		&		&		&		&		&		&	0.0005	&	-0.0044	&	0.0054	\\	\hline
   $\phi$	&	0.0437	&	0.0284	&	0.0726	&	0.0075	&	0.0029	&	0.0252	&	0.0106	&	0.0030	&	0.1431	\\	
$\sigma_m$ 	&	0.6382	&	0.5760	&	0.7062	&	0.5057	&	0.3739	&	0.7265	&	0.4108	&	0.2859	&	0.5366	\\	\hline

\end{tabular}}
\end{center}

\end{table}


\begin{table}[hptb]
\caption{Model for examining human activity on NO$_2$ concentration: the difference between concentrations measured at urban locations and predictions based on global variables. Models are fit with (i)  intercept term, (ii)  urban covariates.}
\label{ch4:result:model2}
\begin{center}
\scalebox{1}{
\begin{tabular}{l|rrr|rrr}

  & \multicolumn{3}{c}{(i)} & \multicolumn{3}{c}{(ii)}   \\

&     Median &   $2.5\%$ &  $97.5\%$
           &     Median &   $2.5\%$ &  $97.5\%$\\ \hline

\hline

Intercept (urban)   	&	3.2940	&	3.2630	&	3.3270	&	0.8369	&	0.4291	&	1.2570	\\	
Major road	&		&		&		&	0.0623	&	0.0036	&	0.1186	\\	
Minor road	&		&		&		&	0.0266	&	-0.0171	&	0.0653	\\	
High density residential	&		&		&		&	0.0041	&	0.0003	&	0.0088	\\	
Low density residential	&		&		&		&	0.0013	&	-0.0025	&	0.0051	\\	
Industrial	&		&		&		&	0.0016	&	-0.0029	&	0.0064	\\	
Transport	&		&		&		&	0.0069	&	-0.0031	&	0.0173	\\	
Sea port 	&		&		&		&	0.0072	&	-0.0129	&	0.0245	\\	
Air port 	&		&		&		&	0.0001	&	-0.0135	&	0.0154	\\	
Construction	&		&		&		&	0.0058	&	-0.0044	&	0.0154	\\	
Urban greenery	&		&		&		&	-0.0022	&	-0.0115	&	0.0058	\\	\hline
\end{tabular}}
\end{center}
\end{table}

\clearpage
\newpage


\end{document}